\documentclass[aps,twocolumn,amsmath,amssymb]{revtex4}

\usepackage{graphicx}

\begin{document}

\title{Berry phase for ferromagnet with fractional spin}
\author{A.~G.~Abanov}
\affiliation{
	    Department of Physics and Astronomy,
            Stony Brook University,
            Stony Brook, NY 11794-3800
}

\author{Ar.~Abanov}
\affiliation{
            Los Alamos National Laboratory,
	    Theoretical division, MS262,
            Los Alamos, NM 87545
}

%\date{\today}

\begin{abstract}
We study the double exchange model on two lattice sites with one
conduction electron in the limit of an infinite Hund's interaction. 
While this simple problem is exactly solvable, we present an
approximate solution which is valid in the limit of large core spins. 
This solution is obtained by integrating out charge degrees of
freedom.  The effective action of two core spins obtained in the
result of such an integration resembles the action of two fractional
spins.  We show that the action obtained via naive gradient expansion
is inconsistent.  However, a ``non-perturbative'' treatment leads to
an extra term in the effective action which fixes this inconsistency. 
The obtained ``Berry phase term'' is geometric in nature.  It arises
from a geometric constraint on a target space imposed by an adiabatic
approximation.
\end{abstract}
\vspace{0.15cm}

\maketitle

%%%%%%%%%%%%%%%%%%%%%%%%%%%%%%%%%%%%%%%%%%%%%%%%%%%%%%%%%%%%%%%%%
\section{Introduction}
%%%%%%%%%%%%%%%%%%%%%%%%%%%%%%%%%%%%%%%%%%%%%%%%%%%%%%%%%%%%%%%%%

A double exchange model was introduced in \cite{Zener-1951,AH-1955} to
describe the motion of conduction electrons in the background of
magnetic ions.  In its simplest form the Hamiltonian for the model can
be written as:
\begin{equation}
       H=-\sum_{\langle ij\rangle}(tc_{i}^{\dagger}c_{j}+h.c.)
       -\sum_{i}J_{H}\vec{s}_{i} \vec{S}_{i}
    \label{eq:de},
\end{equation}
where $c_{i}$ is the electron annihilation operator on the site $i$ of
the lattice, $\vec{s}_{i}=\frac{1}{2}c_{i}^{\dagger} \vec{ \sigma
}c_{i}$ is a spin $1/2$ of an electron \footnote{Here and in the
following we use brief ``matrix'' notations so that, e.g., this
expression means actually $\frac{1}{2}c_{i\alpha}^{\dagger}
\vec{\sigma}_{\alpha\beta}c_{i\beta}$ with implicit summation over
repeated spin indices $\alpha,\beta=1,2$.} at site $i$, $\vec{S}_{i}$
is a core spin.  The first term of (\ref{eq:de}) describes the hopping
of conduction electrons between nearest neighbor sites $i$ and $j$ of
the lattice.  The hopping amplitude is $t$.  The second term is a
ferromagnetic Hund's interaction between the core spin and the spin of
the conduction electron at the same lattice site.  To specify the
model completely we have to fix the total number of conduction
electrons in the sample.  We denote filling factor
$x=\sum_{i}c_{i}^{\dagger}c_{i}/N$, where $N$ is a total number of
sites of the lattice.  In this paper we are interested
\cite{double_exchange} in the limit of a very strong Hund's
interaction $J_{H}/t\to \infty$ and of very large core spins $S\gg 1$. 
In (\ref{eq:de}) we did not include any (bare) core spin interactions,
e.g., due to superexchange.  The only interaction between core spins
in the model (\ref{eq:de}) occurs through the exchange by conduction
electrons.

Because of an infinite Hund's coupling
$J_{H}$ an electron spin is always aligned with a core spin on any
given lattice site.  Therefore, the effective hopping amplitude of an electron
between two nearest neighbors is the largest when corresponding core spins
are parallel.  In this case an electron can hop between lattice sites without
changing its spin orientation.  One concludes that conduction
electrons induce a ferromagnetic interaction between core spins
\cite{AH-1955,deGennes-1960}.

In the limit of large core spins $S\gg 1$ we consider one can use 
an adiabatic approximation and
think of core spins as of slow variables so that the conduction
electrons (fast variables) are always in the ground state in the
background of slowly moving core spins.  
It is tempting to average over the motion of conduction electrons (to
integrate conduction electrons out) and obtain an effective spin model
describing the dynamics of core spins (and electron spins) induced by
charge degrees of freedom \footnote{One should of course remember that
this is a dangerous thing to do because charge degrees of freedom are
not gapped in infinite system and one should expect some nonlocal
spin-spin interactions induced in the process of such an
integration.}.  In an almost ferromagnetic state the average spin per
lattice site is $S+\frac{1}{2}x$ corresponding to the sum of a core
spin and an average electron spin.  Therefore, heuristically, one
arrives to the model of ferromagnetically interacting {\em fractional}
spins.  However, this immediately raises the question of what one
means by a fractional spin.  Although the average spin per lattice
site is fractional, the total spin operator on a given site may take
only values $S$ (no electron on the site) or $S+1/2$ (there is an
electron on the site).  Moreover, a fractional value of $2S$ is
incompatible with quantum mechanics of spins (see section
\ref{sec:singlespin}).  Resolving this apparent controversy is a main
goal of this paper.

The paper is organized in the following way.  The aim of the section
\ref{sec:singlespin} is mainly pedagogical.  We establish notations
and remind the reader how to write the path integral for a single
spin.  We show how the ambiguity of the Berry phase for a single spin
requires $2S$ be an integer.  For completeness we derive the Berry
phase for a single spin in Appendix \ref{app:pathint} using a
fermionic determinant representation for the action.  The reader
familiar with the subject can start from the section \ref{sec:deInfJ}
where we consider the double exchange model (\ref{eq:de}) in the limit
$J_{H}/t\to \infty$ and $S\gg 1$.  In section \ref{sec:de2} we
formulate a double exchange model on two core spins and obtain our
main result -- an effective action for core spins induced by a single
conduction electron.  This action (namely, its kinetic part) answers
the question of how to write Berry phase for a system of (two)
fractional spins.  Appendix \ref{app:diagrams} contains more accurate
calculation of the effective action which uses the method of gradient
expansion.  We discuss the topology behind Berry phase for fractional
spins in section \ref{sec:tabp} and conclude in section
\ref{sec:concl}.

%%%%%%%%%%%%%%%%%%%%%%%%%%%%%%%%%%%%%%%%%%%%%%%%%%%%%%%%%%%%%%%%%
\section{Path integral for single spin}
    \label{sec:singlespin}
%%%%%%%%%%%%%%%%%%%%%%%%%%%%%%%%%%%%%%%%%%%%%%%%%%%%%%%%%%%%%%%%%
                                                                 
In this section we remind the reader how to write down the path integral
for a single quantum spin and discuss topological reasons for spin
quantization.

Quantum spin $\vec{S}$ is defined as  a 
representation of dimension
$2S+1$ of an $SU(2)$ Lie algebra with commutation relations
\begin{eqnarray}
          \left[S^{+},S^{-}\right] &=& 2S^{z},
     \nonumber \\
          \left[S^{z},S^{\pm}\right] &=& \pm S^{\pm},
     \label{eq:su2}
\end{eqnarray}
where as usual $S^{\pm}=S^{x}\pm iS^{y}$. An $SU(2)$
Lie algebra has
irreducible finite dimensional representations of dimension
$2S+1$ labelled by spin $S$ taking integer or half-integer values
so that $\vec{S}^{2}=S(S+1)$.  We would like to construct
the classical action for the time dependent
unit vector $\vec{n}(t)$ which upon
quantization by path integral reproduces (\ref{eq:su2}).
One could start directly from (\ref{eq:su2}) and construct the
path integral using coherent states corresponding to an $SU(2)$
group (see e.g., Ref.\cite{Fradkin-1991}).
In Appendix \ref{app:pathint} we give an
alternative derivation of this path integral using the 
fermionic determinant representation.

The result is given by partition function
\begin{eqnarray}
          Z &=& \int D\vec{n}\;  e^{-W_{E}[\vec{n}]},
     \label{eq:partition}
\end{eqnarray}
where
\begin{eqnarray}
       W_{E}[\vec{n}] &=&
       i (2S) 2\pi \int_{\cal B} d^{2}x\,
       \frac{1}{8\pi}\epsilon^{abc}\epsilon_{\mu\nu}
       n^{a}\partial_{\mu}n^{b}\partial_{\nu}n^{c}.
    \label{eq:WZNW}
\end{eqnarray}
Let us explain what (\ref{eq:partition},\ref{eq:WZNW}) mean. First of
all we used an Euclidean time representation replacing real physical
time $t$ by an imaginary time $\tau =it$. The change to an
imaginary time changes quantum amplitude $e^{iW}$ used
in path integral into a
``Boltzmann'' factor $e^{-W_{E}}$.
We impose periodic boundary conditions (in imaginary time)
on $\vec{n}(\tau)$ so that $\vec{n}(\beta)=\vec{n}(0)$ and
$\beta$ is a maximal span of an imaginary time.  With these
boundary conditions one can think of time as of a
one-dimensional circle with identification of point
$\tau=\beta$ with $\tau=0$.  Physical results will be
obtained in the limit $\beta\to \infty$ after an analytical
continuation back to the real time $t=-i\tau$.
Imaginary time
representation is slightly more convenient for calculations and for
a discussion of topological aspects of spin quantization.

Secondly, the two-dimensional integration in (\ref{eq:WZNW}) is
performed over some closed
domain ${\cal B}$ of two dimensional plane such that the
boundary of this domain is identified with an imaginary time.
The values of
$\vec{n}$ field in the area ${\cal B}$ are defined as
an extension from the physical trajectory $\vec{n}(\tau)$ so
that $\vec{n}(\tau,u)$ is a smooth function of
$\tau\in[0,\beta]$ and auxiliary variable $u\in [0,1]$ such
that $\vec{n}(\tau,u=0)=(0,0,1)$ and
$\vec{n}(\tau,u=1)=\vec{n}(\tau)$.
We shall refer to the action (\ref{eq:WZNW}) as to a Berry 
phase \cite{Berry-1984} for 
a spin (see Appendix \ref{app:pathint}).
Although the action (\ref{eq:WZNW}) is given in terms of
two-dimensional integral over auxiliary space, the variation of the
integrand is a full derivative. Therefore, the variation of
Berry phase (\ref{eq:WZNW}) is given by a one-dimensional integral over
the ``physical'' (imaginary) time as
\begin{eqnarray}
       \delta W_{E} &=&
       iS\int_{0}^{\beta} d\tau \,
       \epsilon^{abc}
       \delta n^{a}n^{b}\partial_{\tau}n^{c}.
  \label{eq:variation}
\end{eqnarray}
We see that the variation
of (\ref{eq:WZNW}) does not depend
on the choice of an extension of $\vec{n}(\tau)$ in the
unphysical domain ${\cal B}$. Using (\ref{eq:variation}) one can show
that the equation $\delta(W_{E}+W_{E}^{h})=0$, where
$W_{E}^{h}=\int_{0}^{\beta}d\tau \, S\vec{h}\vec{n}$ is a coupling to an
external magnetic field gives a classical equation of motion
$i\partial_{\tau}\vec{n}=[\vec{h}\times\vec{n}]$. The
latter is the equation of precession of spin in magnetic field $\vec{h}$.

Partition function (\ref{eq:partition}) with the action
(\ref{eq:WZNW})
solve the problem
of path integral representation for the algebra
(\ref{eq:su2}). Namely, the commutation relations (\ref{eq:su2}) can
be obtained as a result of quantization \cite{Fradkin-1991} 
of ``classical action'' (\ref{eq:WZNW}).

\subsection{Berry phase and quantization of spin}

\begin{figure}
   \includegraphics[width=\columnwidth]{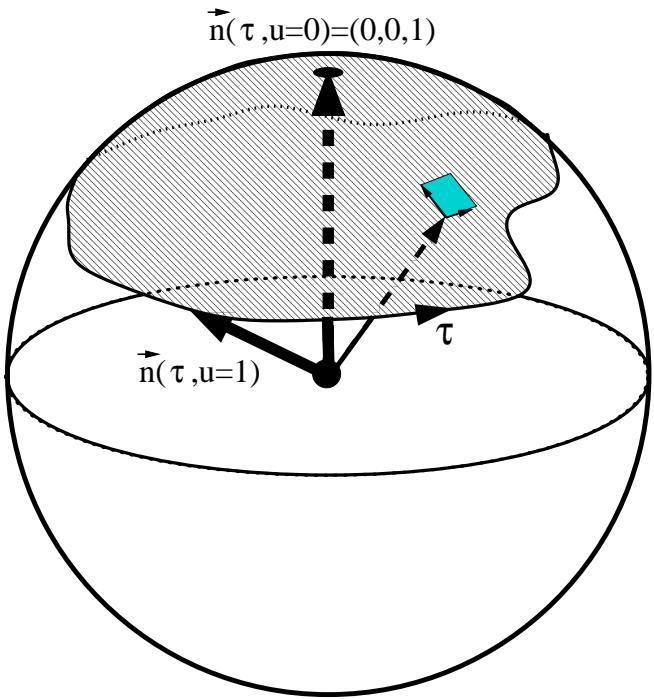}
   \caption{\label{fig:WZNW} The unit vector $\vec{n}(\tau)$ draws a 
   closed line on the surface of a sphere with unit radius during its 
   motion in imaginary time. Berry phase is proportional 
   to the solid angle
   (shaded region) swept by the vector $\vec{n}(\tau)$. One can 
   calculate this solid angle by extending $\vec{n}$ into the 
   two-dimensional domain ${\cal B}$ as $\vec{n}(u,\tau)$ and 
   calculating (\protect\ref{eq:WZNW}).}
\end{figure}

We have already checked that an infinitesimal variation of
Berry phase $\delta W_{E}[\vec{n}]$ does not depend on the extension of
$\vec{n}(\tau)$ into the auxiliary domain ${\cal B}$. How
about $W_{E}[\vec{n}]$ itself? One can easily show (see Appendix 
\ref{app:pathint}) that the action
(\ref{eq:WZNW}) is equal to $i(2S)
\Omega/2$ where $\Omega$ is a solid angle
swept by a unit vector $\vec{n}$ during its motion (see Fig.\
\ref{fig:WZNW}).  This angle is not uniquely
defined: one can always add or subtract any angle
which is a multiple of a full solid angle $4\pi$.

Therefore, $W_{E}[\vec{n}]$ is not unambiguously defined and is
a multi-valued functional of $\vec{n}(\tau)$. Multi-valued functionals
of this type were studied in both mathematics and quantum field 
theory (see e.g., \cite{currentalgebras}) and are often 
referred to as WZNW term  ---
Wess-Zumino-Novikov-Witten term. 
Although $W_{E}(\vec{n})$ is a multiple-valued functional, 
in partition function (\ref{eq:partition}) one
needs $e^{-W_{E}[\vec{n}]}$ which is a single-valued functional
of $\vec{n}(\tau)$ under the condition that the coefficient
$2S$ is an integer. Then $e^{-\delta W_{E}}=e^{-i(2S)2\pi k}=1$ and
the Boltzmann weights $e^{-W_{E}(\vec{n})}$ is defined unambiguously by
physical configurations $\vec{n}(\tau)$. One can think of this
constraint on the value of $2S$ as of the {\em reason} for
spin to be an integer or a half-integer. Having non-integer $2S$
in (\ref{eq:WZNW}) amounts to having {\em inconsistent}
theory.

\subsection{$CP^{1}$ representation of Berry phase and topology}

There is a one more representation of Berry phase which is 
particularly convenient for the purposes of this paper. It is
a so-called $CP^{1}$ representation. We write
$\vec{n}=z^{\dagger}\vec{\sigma}z$, where
$z=(z_{1},z_{2})^{t}$ is a complex two-component vector satisfying
the constraint $z^{\dagger}z=|z_{1}|^{2}+|z_{2}|^{2}=1$. The latter
reproduces $\vec{n}^{2}=1$. 

In terms of $z$-variable we have for
Berry phase
\begin{eqnarray}
       W_{E}[\vec{n}] &=& -i (2S) \int_{0}^{\beta} d\tau \,
       z^{\dagger}i\partial_{\tau}z
    \label{eq:WZNWCP1}
\end{eqnarray}
which is apparently independent from the auxiliary coordinate
$u$.  This ``simplification'' is, however, deceptive
because the equation $\vec{n}=z^{\dagger}\vec{\sigma}z$ does
not define $z$ unambiguously.  
One can always change $z(\tau)\to e^{i\alpha(\tau)}z(\tau)$
with any function $\alpha(\tau)$ satisfying
$e^{i\alpha(\beta)}=e^{i\alpha(0)}$. The latter is necessary to 
conserve periodic boundary 
conditions $z(\beta)=z(0)$. This ``gauge'' transformation
obviously does not change the vector $\vec{n}$. The change of
WZNW action (\ref{eq:WZNWCP1}) 
under this transformation is $\delta W_{E} =
i(2S)[\alpha(\beta)-\alpha(0)]= i (2S)2\pi k$ where $k$
is any integer. We arrive to the same result: the action
$W_{E}$ itself is multivalued while $e^{-W_{E}}$ is single-valued
under {\it the condition} that $2S$ is an {\it integer number}!

Let us now briefly discuss topological aspects of Berry phase
(WZNW term) for a single spin (\ref{eq:WZNW}).
The classical unit vector $\vec{n}$ from (\ref{eq:WZNW}) takes values
on a unit two-dimensional sphere $S^{2}$ -- the target space of the
problem. There exists nontrivial expression (the integrand of
(\ref{eq:WZNW})) depending on vector $\vec{n}$ and
its derivatives which is (i) local (ii) being integrated over
two-dimensional sphere gives integer values for all possible smooth
configurations of $\vec{n}$ defined on $S^{2}$. Mathematically we say
that there exists a closed but non-exact 2-form on $S^{2}$ (target
space) or even more formally that the second cohomology group of $S^{2}$ is
$H_{2}(S^{2})=Z$. This allows one to write down the topological term
(\ref{eq:WZNW}) where an integration of the mentioned 2-form is performed
over a domain ${\cal B}$ with physical time as the boundary of this disk. 
The special properties mentioned above guarantee that (\ref{eq:WZNW})
unambiguously defines the path integral (\ref{eq:partition}).

%%%%%%%%%%%%%%%%%%%%%%%%%%%%%%%%%%%%%%%%%%%%%%%%%%%%%%%%%%%%%%%%%
\section{Double exchange model in the limit of infinite $J_{H}$}
    \label{sec:deInfJ}
%%%%%%%%%%%%%%%%%%%%%%%%%%%%%%%%%%%%%%%%%%%%%%%%%%%%%%%%%%%%%%%%%

Let us now consider the double exchange model (\ref{eq:de}) and take
a limit of infinite Hund's constant
$J_{H}/t\to\infty$. We will do it in the Lagrangian formulation. First we
use the WZNW action (\ref{eq:WZNWCP1}) to write the Lagrangian
(in this section we use real time representation)
of the double exchange model
corresponding to the Hamiltonian (\ref{eq:de})
\begin{equation}
       L = \sum_{i}\left[c_{i}^{\dagger}(i\partial_{t}-\mu)c_{i}
       + 2S z_{i}^{\dagger}i\partial_{t}z_{i}\right] - H.
       \label{eq:deLagr}
\end{equation}
Here $\mu$ is a chemical potential which is chosen 
to guarantee the correct
filling factor of electrons 
and $z_{i}=(z_{i1},z_{i2})^{t}$ is a complex vector
corresponding to the core spin $S_{i}$ at the site $i$ of
the lattice. We will make the following change of variables
in the Lagrangian $c_{i}=\psi_{i}z_{i} +\tilde\psi_{i}\tilde
z_{i}$. Here $\psi_{i},\tilde\psi_{i}$ are spinless
fermions, $z_{i}$ is  a core spin variable, and $\tilde
z_{i}$ is a complex vector orthogonal to $z_{i}$, namely,
$\tilde z_{i}=(z_{i2}^{\ast},-z_{i1}^{\ast})^{t}$.
Physically, $\psi_{i}$ is a component of an
electron on the site $i$ with a spin parallel to the core spin
at the same site and $\tilde\psi_{i}$ is the one with a spin 
antiparallel to the
core spin. Taking limit $J_{H}\to \infty$ corresponds to
neglecting $\tilde\psi$ component of an electron.
We substitute $c_{i}=\psi_{i}z_{i}$ into (\ref{eq:deLagr})
and omitting constant terms obtain
\begin{eqnarray}
       L &=& \sum_{i}
       \psi_{i}^{\dagger}(i\partial_{t}-\mu)\psi_{i}
       +\sum_{\langle ij\rangle} \left[
       \psi_{i}^{\dagger}(tz_{i}^{\dagger}z_{j})\psi_{j}
       +h.c.\right]
  \nonumber \\
       &+& \sum_{i}(2S+\psi_{i}^{\dagger}\psi_{i})
       z_{i}^{\dagger}i\partial_{t}z_{i}.
  \label{eq:deLagrProj}
\end{eqnarray}

The first term of (\ref{eq:deLagrProj}) is a kinetic term for 
projected electrons. Let us take a close look at the second and the 
third terms.  We assume that the motion of spins is much slower 
then the motion of electrons. 
This is justified in the limit of large core spins
$S\gg 1$. Then one can think of electrons as  moving
in the background of almost static $z$'s.  The second term
in (\ref{eq:deLagrProj}) describes the hopping of these
electrons with the effective hopping amplitude $t^{eff}_{ij}
= tz_{i}^{\dagger}z_{j}$.  This amplitude is complex number and
has both the magnitude and the phase.

The magnitude of the effective hopping amplitude is given by
$|t^{eff}_{ij}|^{2}\equiv\Delta^{2}=t^{2}(z_{i}^{\dagger}z_{j})
(z_{j}^{\dagger}z_{i}) = t^{2} \frac{1+\vec{n}_{i}
\vec{n}_{j}}{2}$ or $\Delta =t\cos\frac{\theta_{ij}}{2}$ where $
\theta_{ij}$ is the angle between two core spins. In this
derivation we used a relation
$\vec{n}_{i}=z_{i}^{\dagger}\vec{\sigma}z_{i}$ and the known
identity on Pauli matrices $\vec{\sigma}_{\alpha\beta}
\vec{\sigma}_{\gamma\delta}
=2\delta_{\alpha\delta}\delta_{\beta\gamma}
-\delta_{\alpha\beta}\delta_{\gamma\delta}$. The
absolute value of the effective hopping is proportional to
the cosine of the half of the angle between core
spins~\cite{AH-1955}. It is
maximal and equal $t$ for parallel core spins and vanishes
for antiparallel spins.

The phase of the effective amplitude $t^{eff}_{ij}$ is not
defined unambiguously because of the gauge freedom (see the previous 
section).  Consider, however, an
electron moving along some closed trajectory in a background
of a smooth configuration of core spins.  Then along the
path we can write $z_{i+1} = z_{i}+\delta z_{i}$,
  where $\delta z_{i}$ is small, and
$z^{\dagger }_{i}z_{i+1} = 1+z^{\dagger }_{i}\delta z_{i}$.
The product of all the effective amplitudes along the path
is $\prod _{i}t^{eff}_{i,i+1} \sim \exp (\sum_{i}\ln
z^{\dagger }_{i}z_{i+1}) \approx \exp(\sum_{i}z^{\dagger
}_{i}\delta z_{i}) = \exp i\sum_{i}(\delta \psi_{i}+\delta
\phi _{i}\sin ^{2}\theta_{i}/2)$, where we use
parameterization $z_{i}^{t}=e^{i\psi_{i}}(\cos \theta_{i}/2,
e^{i\phi_{i}}\sin\theta_{i}/2 )$ of $z_{i}$ in terms of
Euler angles $\theta_{i},\phi_{i}$, and $\psi_{i}$.  Along
the closed path $\sum_{i}\delta \psi_{i} = 0$, so only the
second term contributes to the phase picked up by an electron.  
It is easy to see that this second term is equal to
the half of the solid angle enclosed by 
vectors $\vec{n}$ along the trajectory of an electron.  We
see that (projected) electrons moving along some closed trajectory
acquire the phase equal to the half of 
solid angle formed by spins along this trajectory. One can say, 
therefore, that they ``feel'' the solid angle
formed by core spins as a flux of magnetic
field. This flux is (modulo $2\pi$) a gauge invariant and well defined 
quantity.

Finally, the last term in (\ref{eq:deLagrProj}) is a
``modified'' kinetic term for core spins.  One might expect
that after the averaging over electrons this term will
describe the dynamics of effective spins of magnitude
$S+\langle \psi^{\dagger }\psi\rangle/2$.  However, having non-integer 
coefficient in front of Berry phase results in an inconsistent theory.

To make this seeming paradox more precise we average over electrons 
accurately by integrating fermions 
\cite{NO-1998} out of (\ref{eq:deLagrProj}).  
We obtain an effective action for core spins in terms of $z$-variables
\begin{eqnarray}
       W &=&
       -i\ln\det\left[\delta_{ij}(i\partial_{t}-\mu
       +z_{i}^{\dagger}i\partial_{t}z_{i})
       +tz_{i}^{\dagger}z_{j}\right]
  \nonumber \\
       &+& \int dt\,
       \sum_{i} 2S
       z_{i}^{\dagger}i\partial_{t}z_{i} .
    \label{eq:deEff}
\end{eqnarray}
Here the expression in the determinant is both an operator
in time and $N\times N$ matrix in lattice sites. The matrix element
$z_{i}^{\dagger}z_{j}$ is present only if $i,j$ are nearest
neighbors. In the adiabatic limit there are two main effects produced
by an integration over fermionic degrees of freedom. 
First, the energy
of the filled (to the filling factor $x$) Fermi sea is added to the
action. This leads to an effective ferromagnetic interaction between
core spins. Second, the correction to the Berry phase of core spins
must reflect the fact that spins of conduction
electrons are effectively added to core spins. The latter correction is
the matter of consideration of this paper.

The variation of (\ref{eq:deEff}) over $z_{i}^{\dagger}i\partial_{t}z_{i}$
immediately gives $(2S+\langle\psi_{i}^{\dagger}\psi_{i}\rangle)$,
where $\langle \psi_{i}^{\dagger}\psi_{i}\rangle$ are some fractional
numbers which in general depend on the configuration of core spins
($z_{i}$'s). This implies the presence of the term
$(2S+\langle\psi_{i}^{\dagger}\psi_{i}\rangle)
z_{i}^{\dagger}i\partial_{t}z_{i}$ in the action (\ref{eq:deEff}) and 
confirms the result we expected from the heuristic ``averaging'' 
over electrons. However, the presence of this term alone renders 
theory inconsistent.

In the next section we calculate the fermionic determinant of the 
type (\ref{eq:deEff}) albeit for a simplified double exchange model on two 
lattice sites. We show that there is an additional term present in 
the effective action for core spins. This additional term will make 
the effective theory consistent as it is shown in section \ref{sec:tabp}.

%%%%%%%%%%%%%%%%%%%%%%%%%%%%%%%%%%%%%%%%%%%%%%%%%%%%%%%%%%%%%%%%%
\section{Double exchange model for two core spins}
    \label{sec:de2}
%%%%%%%%%%%%%%%%%%%%%%%%%%%%%%%%%%%%%%%%%%%%%%%%%%%%%%%%%%%%%%%%%

The paradox of having fractional spin in the effective action for a 
double exchange model is present already for a much simpler double 
exchange model on two lattice sites with a single conduction electron. 
In addition the electron spectrum in such model is discrete which 
makes integration over fermions in adiabatic approximation to be a 
well-defined procedure. Therefore, in this section we 
consider the double exchange model on two
lattice sites with one conduction electron and show how to resolve
the issue of fractional spin on this simple example.

The projected ($J_{H}\to \infty$) double exchange model
(\ref{eq:deLagrProj}) on two lattice sites (in imaginary time)
\begin{eqnarray}
       L_{E} &=& \sum_{i=1,2}\left[
       \psi_{i}^{\dagger}(\partial_{\tau}+i\mu)\psi_{i}
       + (2S+\psi_{i}^{\dagger}\psi_{i})
       z_{i}^{\dagger}\partial_{\tau}z_{i}\right]
  \nonumber \\
       &-& t\left[
       \psi_{1}^{\dagger}(z_{1}^{\dagger}z_{2})\psi_{2}
       +h.c.\right] ,
  \label{eq:deLagrProj2}
\end{eqnarray}
where $W_{E}=\int_{0}^{\beta}d\tau\, L_{E}$ is a Euclidean action.
Chemical potential $\mu$ must be chosen to guarantee that there is
exactly one conduction electron in
the system, i.e., $\sum_{i=1,2}\psi_{i}^{\dagger}\psi_{i}=1$. One can 
repeat the heuristic argument of the previous section replacing  
$\psi_{i}^{\dagger}\psi_{i}$ in the second term of (\ref{eq:deLagrProj2}) 
by its average value $1/2$. One obtains two effective spins of 
magnitude $S+1/4$ instead of bare core spins. Again, we have the 
problem of having the fractional coefficient in the 
Berry phase term. 

We rewrite the Lagrangian (\ref{eq:deLagrProj2}) as
\begin{equation}
       L_{\rm E} = \sum_{i=1,2}\left[
       2S
       z_{i}^{\dagger}\partial_{\tau}z_{i}\right]
       +\Psi^{\dagger}(\partial_{\tau}+i\mu+\hat{A})\Psi .
       \label{eq:proj2}
\end{equation}
Here we defined
$\Psi^{\dagger}=(\psi_{1}^{\dagger},\psi_{2}^{\dagger})$ and
$2\times 2$ matrix $\hat A$
\begin{equation}
     \hat{A} = \left(
     \begin{array}{cc}
           z_{1}^{\dagger}\partial_{\tau}z_{1}
	  & -\Delta e^{-ia} \\
           -\Delta e^{ia}
	  & z_{2}^{\dagger}\partial_{\tau}z_{2}
     \end{array}\right),
  \label{eq:A2}
\end{equation}
where we introduced explicitly the phase and the magnitude of an 
effective hopping amplitude 
\begin{equation}\label{eq:notation}
tz_{1}^{\dagger}z_{2}=\Delta e^{-ia}.
\end{equation}
The absolute
value of an effective hopping amplitude $\Delta=t\cos\theta/2$, where
$\theta$ is an angle between core spins $\vec{S}_{1}$ and
$\vec{S}_{2}$.  It is equal to zero only for strictly antiparallel
core spins $\theta=\pi$.  For fixed directions of core spins the
matrix $\hat A$ becomes constant matrix with eigenvalues $\pm \Delta$
corresponding to the two-level spectrum of an electron hopping between
two core sites in double exchange model with infinite $J_{H}$. 
Two-level spectrum is symmetric and one can choose the chemical potential
$\mu=0$ so that there is only one filled energy level at any given
moment of time. Then the
spinless fermion $\psi$ (electron with projected spin) occupies the
lowest energy level $E=-\Delta$ which is minimal ($E_{\rm min}=-t$)
when core spins are parallel.  Changing the angle between core spins
requires the energy of the order of $t$ while we are interested in low
energy dynamics with typical values of energy $\sim t/S$.  Therefore,
we assume that $\Delta$ does not vanish and is of the order of $t$.

The Grassman variable $c(\tau)$
corresponding to the conduction electron must satisfy antiperiodic
boundary conditions in imaginary time $\tau$: $c_{i}(\beta)=-c(0)$.
We represent $c_{i}(\tau ) =\psi_{i}(\tau)z_{i}(\tau)$
as a product of spinless fermion $\psi_{i}(\tau)$ and spin $1/2$
boson $z_{i}(\tau )$, which after projection $J_{H}\to \infty$
becomes $z$-boson corresponding to core spins. The boundary
conditions for $\psi_{i}(\beta)=-\psi_{i}(0) $ and $z_{i}(\beta) =
z_{i}(0)$ are antiperiodic and periodic
respectively. Then we have for $\Psi(\tau)$ and
$a(\tau)$ (see (\ref{eq:proj2},\ref{eq:notation}))
\begin{eqnarray}
     \Psi (\beta) &=& -\Psi (0),
  \label{eq:boundaryPsi} \\
     a(\beta) &=& a(0) +2\pi k,
  \label{eq:boundaryA}
\end{eqnarray}
where $k$ is an integer -- the winding number of the phase $a$ in
imaginary time $2\pi k=\int_{0}^{\beta}d\tau \, \partial_{\tau}a $.

The effective action of core spins is
given by
\begin{equation}
     W_{E}^{eff} = \int_{0}^{\beta}\sum_{i=1,2}\left[
       2S z_{i}^{\dagger}\partial_{\tau}z_{i}\right] \,d\tau
       +W_{E}^{ind},
  \label{eq:ea}
\end{equation}
where the effective action of core spins induced by fermions
\begin{equation}
     W_{E}^{ind} = -\,{\rm{Tr}\,}\ln(\partial_{\tau }+ \hat {A}).
  \label{eq:ft2}
\end{equation}
Here the operator $\partial_{\tau}+\hat{A}$ is linear differential
operator with explicit dependence on time $\tau$
through the $2\times 2$ matrix $\hat{A}$ and we put $\mu=0$
corresponding to having only one conduction electron in the system.
To calculate the functional trace (\ref{eq:ft2}) we find the
eigenvalues of this operator by solving
differential equation
\begin{equation}
  \label{eq:equation1}
     (\partial_{\tau } +\hat {A})\Psi(\tau)=\lambda \Psi(\tau )
\end{equation}
with boundary conditions (\ref{eq:boundaryPsi},\ref{eq:boundaryA})
and calculate the sum of logarithms of all eigenvalues $\lambda$.
Matrix (\ref{eq:A2}) can be rewritten as
\begin{equation}
     \hat{A} = \frac{i}{2}A_{0}
     -e^{-\frac{i}{2}a\sigma^{3}}\Delta\sigma^{1} e^{\frac{i}{2}a\sigma^{3}}
     +ia_{0}\sigma^{3},
  \label{eq:A20}
\end{equation}
where
\begin{eqnarray}
     iA_{0} &=& z_{1}^{\dagger }\partial_{\tau } z_{1}+
     z_{2}^{\dagger }\partial_{\tau } z_{2} ,
  \nonumber \\
     ia_{0} &=& \frac{1}{2}\left(z_{1}^{\dagger }\partial_{\tau } z_{1}-
     z_{2}^{\dagger }\partial_{\tau } z_{2} \right) .
  \nonumber
\end{eqnarray}
We would like to solve (\ref{eq:equation1}) with the matrix $\hat{A}$
from (\ref{eq:A20}) in an adiabatic approximation, e.g., assuming that all
time derivatives are much smaller than the energy scale $\Delta\sim t$.
First we ``unwind'' wavefunction $\Psi$ by
a unitary transformation
\begin{equation}
  \label{eq:transformation}
     \Psi = \exp\left\{-\frac{i}{2}\int_{0}^{\tau}d\tau\,
     (A_{0}+\dot{a}\sigma^{3})\right\}\chi,
\end{equation}
where $\dot{a}$ means the derivative of $a$ over $\tau $. 
Then, instead of
(\ref{eq:equation1}) the function $\chi$
satisfies
\begin{equation}
     (\hat{L}_{0} + \hat{V})\chi = \lambda \chi,
  \label{eq:ptequation}
\end{equation}
where we defined ``zero order'' operator $\hat{L}_{0}$ and
perturbation $\hat{V}$ which is proportional to time derivatives of
core spins as
\begin{eqnarray}
     \hat{L}_{0} &=& \partial_{\tau} - \Delta \sigma^{1},
  \nonumber \\
     \hat{V} &=& i\tilde{a}_{0}\sigma^{3},
  \nonumber \\
     \tilde{a}_{0} &=& a_{0}-\frac{1}{2}\dot{a} .
  \nonumber
\end{eqnarray}
Now we can solve (\ref{eq:ptequation}) using the perturbation theory
in $\hat{V}$. The boundary conditions of $\chi$ are given by
(\ref{eq:boundaryPsi}) and (\ref{eq:transformation}) as
\[
\chi (\beta) = -e^{i\gamma}\chi (0),
\]
where Berry phase
\begin{equation}
     \gamma =\frac{1}{2}\int_{0}^{\beta}(A_{0}+\dot{a})\,d\tau,
  \nonumber
\end{equation}
where we used that $\int _{0}^{\beta }\dot{a}d\tau =2\pi k$ 
($k$ is an integer, see (\ref{eq:boundaryA})) and that 
$e^{i\pi k\sigma ^{3}}=e^{i\pi k}$.

We are going to find the eigenvalues of (\ref{eq:ptequation}) using the
perturbation theory in $\hat{V}$. This is justified in an adiabatic limit
when $\tilde{a}_{0}$ proportional to the rate of change of
core spins is much smaller than the typical fermionic energy
$\Delta$ or in other words when the conduction electron is the 
fast degree of freedom compared to core spins.

The solutions of the zero order equation
\[
     \hat{L}_{0}\chi^{(0)}=\lambda^{(0)}\chi^{(0)}
\]
are labelled by an integer number $n$ and by plus or minus sign
\begin{eqnarray}
     |\chi^{(0)}_{n\pm}\rangle &=& \frac{1}{\sqrt{2\beta }}
     e^{\mp\int _{0}^{\tau }\Delta\,d\tau +\lambda^{(0)}_{n\pm}\tau}
     \left(\begin{array}{c}
     1\\
     \pm 1
     \end{array} \right)
  \nonumber
\end{eqnarray}
with eigenvalues
\begin{eqnarray}
     \lambda^{(0)}_{n\pm} &=& \pm\bar{\Delta}
     +\frac{i\pi }{\beta }(2n+1)  +\frac{i}{\beta}\gamma  ,
  \label{eq:zeroOrder}
\end{eqnarray}
where
\[
     \bar{\Delta} = \frac{1}{\beta}
     \int_{0}^{\beta}\Delta \,d\tau.
\]
There are no corrections to these eigenvalues in the first order of
perturbation theory in $\hat{V}$. Therefore, up to the second order of
perturbation theory we obtain
\begin{eqnarray}
     \lambda_{n\pm} &=& -\mu \pm\bar{\Delta}
     +\frac{i\pi }{\beta }\left(2n+1+\frac{\gamma}{\pi}\right) ,
  \label{eq:firstOrder}
\end{eqnarray}
where  we have
restored the chemical potential $\mu$ which should be  
$\mu=0$ for the problem with a single conduction electron.

The effective action (\ref{eq:ft2}) is given by
\begin{equation}
     W_{E}^{ind} = -\sum_{n=-\infty}^{+\infty}\left[
     \ln\lambda_{n+}+ \ln\lambda_{n-}\right].
  \label{eq:ft3}
\end{equation}
To calculate the obviously divergent sum in (\ref{eq:ft3}) we
regularize it  by subtracting the effective action for the system
without conduction electrons. That is we calculate the difference
between (\ref{eq:ft3}) calculated at $\mu=0$ and the one calculated
at $\mu\to -\infty$.
\begin{eqnarray}
     W_{E}^{ind} &=& -\sum_{n=-\infty}^{+\infty}\left[
     \ln\frac{\lambda_{n+}(\mu=0)}{\lambda_{n+}(\mu\to -\infty)} \right.
  \nonumber \\
     &+& \left.
     \ln\frac{\lambda_{n-}(\mu=0)}{\lambda_{n-}(\mu\to -\infty)}
     \right].
  \label{eq:ft4}
\end{eqnarray}
This sum is still logarithmically divergent but one can show that the
divergent part is a constant not depending on physical parameters
$\Delta $ and $\gamma$. Calculating the sum in (\ref{eq:ft4}) we obtain
\begin{widetext}
\begin{eqnarray}
     W_{ind} &=& -\lim_{\mu\to-\infty}
     \ln\frac{
     \cosh\left(\frac{\beta\bar{\Delta}}{2}+i\frac{\gamma}{2}\right)
     \cosh\left(\frac{\beta\bar{\Delta}}{2}-i\frac{\gamma}{2}\right)}
     {\cosh\left(\frac{\beta(\bar{\Delta}-\mu)}{2}+i\frac{\gamma}{2}\right)
     \cosh\left(\frac{\beta(\bar{\Delta}+\mu)}{2}-i\frac{\gamma}{2}\right)}.
  \label{eq:ft5}
\end{eqnarray}
\end{widetext}
In the limit $\mu\to -\infty$ we obtain
\begin{eqnarray}
     W_{ind} &=& i\gamma
     -\ln\cosh\left(\frac{\beta\bar{\Delta}}{2}
     -i\frac{\gamma}{2}\right)
  \nonumber \\
     &-& \ln\cosh\left(\frac{\beta\bar{\Delta}}{2}
     +i\frac{\gamma}{2}\right).
  \label{eq:ft6}
\end{eqnarray}
This is the final result for the effective action calculated at
finite ``temperature'' $\beta$.  Notice that the only imaginary part
of a Euclidean effective action is the Berry phase $\gamma$.  We have
obtained  this phase 
in the first order of perturbation theory.
However, the same expression
is also valid in all orders of perturbation theory.  It
is easy to show for our simple model because the higher orders of
perturbation theory can only give corrections to the real part of
(\ref{eq:firstOrder}) which does not contribute to the imaginary part
of an Euclidean action.  We consider the higher orders of
perturbation theory using the regular gradient expansion in
Appendix \ref{app:diagrams}.  In fact, the imaginary part of a Euclidean
action has a topological nature and, therefore, can not be changed by small
perturbations (see the discussion in section \ref{sec:tabp}).

In the ``zero temperature'' limit $\beta\Delta\gg 1$  we have
\[
W_{E}^{ind} =
i\gamma -\beta\bar{\Delta}
\]
or explicitly
\begin{eqnarray}
     W_{E}^{ind} &=&  \frac{i}{2}\int_{0}^{\beta}(A_{0}+\dot{a})\,d\tau
     -\int_{0}^{\beta}\Delta\,d\tau .
  \label{eq:ft7}
\end{eqnarray}
The latter result can of course be obtained just by replacing the
discrete sum over $n$ in (\ref{eq:ft4}) by an integral.
For the full effective action (\ref{eq:ea}) we obtain
\begin{eqnarray}
     W_{E}^{eff} &=& W_{E}^{B} +W_{E}^{dyn},
  \label{eq:fea} \\
     W_{E}^{B} &=&  \int_{0}^{\beta}d\tau\,\left[\sum_{i=1,2}
       \left(2S+\frac{1}{2}\right)
       z_{i}^{\dagger}\partial_{\tau}z_{i} +\frac{i}{2}\dot{a} \right],
  \label{eq:kin} \\
     W_{E}^{dyn} &=&-\int_{0}^{\beta}\Delta\,d\tau
     +\ldots.
  \label{eq:dyn}
\end{eqnarray}

The first term of the dynamic part of an effective action $W_{E}^{dyn}$ 
is just the time integral of the energy of the ground state of the
conduction electron in a fixed instantaneous configuration of core
spins. The dots denote the higher order corrections
to (\ref{eq:dyn}). These corrections are small if the background core 
spins are moving adiabatically and they are
diverging if derivatives $\dot{\vec{n}}_{1}, \dot{\vec{n}}_{2}$ are
bigger than $\Delta$. For such processes an adiabatic approximation
fails. In particular, if core spins have opposite directions
$\vec{n}_{1}=-\vec{n}_{2}$ we have
$\Delta=0$ and an adiabatic approximation is not applicable at
any rate of the change of core spins.

%%%%%%%%%%%%%%%%%%%%%%%%%%%%%%%%%%%%%%%%%%%%%%%%%%%%%%%%%%%%%%%%%
\section{Berry phase for fractional spins}
    \label{sec:tabp}
%%%%%%%%%%%%%%%%%%%%%%%%%%%%%%%%%%%%%%%%%%%%%%%%%%%%%%%%%%%%%%%%%

\subsection{Consistency}

The Berry phase term (\ref{eq:kin}) is
the main result of this paper. We see that the the correction to
the Berry phase of bare core spins induced by the conduction 
electron is twofold.
First, as we expected 
averaging over fermions in (\ref{eq:deLagrProj2})
there is a change of the coefficient in Berry phases of core spins
from $2S$ to $2S+1/2$. In addition, there is a
correction $\frac{i}{2}\int_{0}^{\beta}d\tau\,\dot{a}= i\pi k$, where
the integer number $k$ is a winding number (\ref{eq:boundaryA})
of the phase $a$ in imaginary time. The latter term is a full time
derivative and, therefore, does not change the classical dynamics of
spins. Classically, the effective ``core'' spins behave as if they had the
value of $S+1/4$!

Let us show now that the full Berry phase (\ref{eq:kin}) is
consistent with quantum mechanics of spins and, therefore, answers the
question of how to write down the Berry phase for two fractional
spins $S+1/4$. As we have seen at the end of the
section \ref{sec:singlespin} the inconsistency occurs if the partition
function is not invariant under gauge transformations of $z$
variables. We perform the gauge
transformation $z_{1,2}\to e^{i\alpha_{1,2}(\tau)}z_{1,2}$ 
independently
for first and second core spins. 
In order not to change the periodic boundary
conditions of $z_{1,2}$ we require
$\Delta\alpha_{1,2}=\alpha_{1,2}(\beta)-\alpha_{1,2}(0) = 2\pi
l_{1,2}$, where $l_{1,2}$ are arbitrary integer numbers. Under this
transformation the three terms  of (\ref{eq:kin}) depending on
$z_{1}$, $z_{2}$, and $a$  change by
$2\pi i l_{1}(2S+1/2)$, $2\pi i l_{2}(2S+1/2)$, and $\pi
i(l_{1}-l_{2})$ respectively. It is easy to see that neither of
these three terms gives an invariant contribution to a partition
function. E.g., for $l_{1}=1$ the change of the first term is $2\pi
i(2S+1/2)$ which gives $-1$ being exponentiated. However, the total
kinetic term (\ref{eq:kin}) is equal to the sum of all these terms and
changes under the gauge transformation by $2\pi i [(2S+1)l_{1}+2Sl_{2}]$
which is a multiple of $2\pi$ and, therefore, does not change the
weight $e^{-W_{E}^{B}}$. This means that neither of three terms
considered makes any sense separately from the 
others \footnote{Authors are grateful to
D. A. Ivanov who attracted
their attention to this fact.}. 
Only their gauge invariant (modulo $2\pi$) sum has a physical
meaning of the combined Berry phase of the system of two core spins
plus one conduction electron.

\subsection{Topology and adiabatic approximation}

We have seen in section \ref{sec:singlespin}  that
the existence of WZNW term for a single unit vector (single
spin) is due to the fact that the target space $S^{2}$ has a nontrivial
second cohomology group $H_{2}(S^{2})=Z$. The addition of WZNW
term to the action with the coefficient $2S$ is consistent only if
this coefficient is quantized (is an integer number). 
For two independent spins the target space becomes
$S^{2}\times S^{2}$ and one obviously can  write two independent
WZNW terms corresponding to $H_{2}(S^{2}\times S^{2})=Z\times Z$.
This is apparently true for two independent spins. However, the result
is different for the conduction electron induced dynamics of core
spins in the double exchange model. Let us discuss why this is so.
To obtain the induced dynamics of
core spins we have excluded the charge degree of
freedom of conduction electron integrating it out in an adiabatic
approximation. This approximation, however, breaks when the the rate
of the change of core spins becomes comparable to the energy scale
$2\Delta$ given by the levels of conduction electron in the
background of time independent core spins. There is a special point
though when core spins are opposite to each other. At this point
$\Delta =0$ and an adiabatic approximation breaks no matter how slow
the motion of core spins is. 

\begin{figure}
\includegraphics[width=\columnwidth]{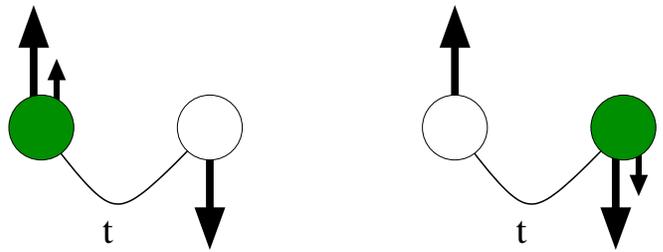}
\caption{\label{fig:oppspins} Two degenerate states with antiparallel 
core spins. An electron with spin aligned along the core spin can not 
hop from one lattice site to another and can be either on the right 
or on the left lattice site.
}
\end{figure}

\begin{figure}
\includegraphics[width=\columnwidth]{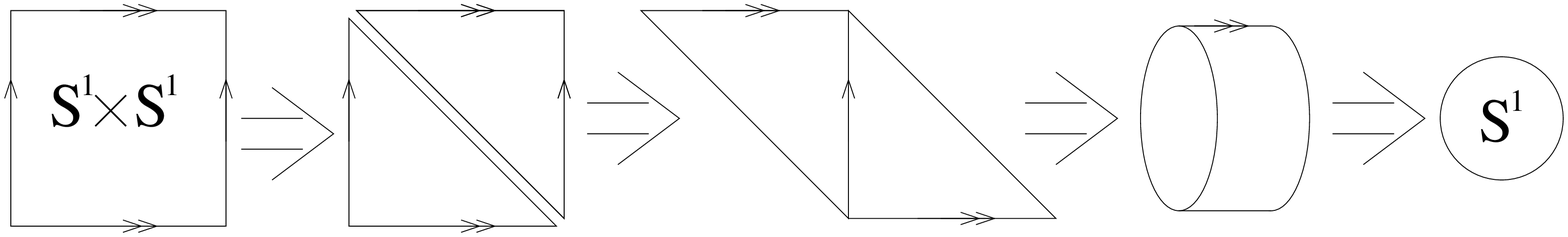}
\caption{\label{fig:cutting} 
The contraction $S^{1}\times S^{1}\backslash S^{1}\to S^{1}$ is shown 
as a one-dimensional illustration of $G\sim S^{2}$. a) 
The torus space $S^{1}\times 
S^{1}$ is represented as a square with the opposite sides 
identified according to the direction of arrows. The diagonal shows 
the points given by $\vec{n}_{1}=-\vec{n}_{2}$. b) The space 
$S^{1}\times S^{1}\backslash S^{1}\to S^{1}$ is obtained by the 
removal of the diagonal. c) One pair of opposite sides is glued. d) 
The second pair is glued making a cylinder. e) The cylinder is 
contracted onto the unit circle.
}
\end{figure}

This is quite obvious from the physical point of
view. Consider two time independent core spins opposite to each other.
The effective hopping amplitude of a conduction electron is $\sim
\cos\theta/2 =0$ and one has two different configurations. The
electron can be either on the left lattice site or on the right one (see
Figure \ref{fig:oppspins}) and the matrix element of the hopping term 
is zero between those states. This means that the ground state of the 
system is 
doubly degenerate. The charge degree of freedom (left or right 
position of the electron)
is essential for this particular configuration of core spins and can
not be integrated out. In our derivation we used an adiabatic
approximation and neglected such configurations. This is justified in
the limit $S\gg 1$ for ferromagnetically interacting core spins.
However, this approximation drastically changes the topology of the
system. Essentially, we prohibited the configurations
$\vec{n}_{1}=-\vec{n}_{2}$. The relevant target space changes from
$S^{2}\times S^{2}$ for two independent spins to $G=S^{2}\times
S^{2}\backslash S^{2}$. The latter expression denotes the direct
product $S^{2}\times S^{2}$ with points $\vec{n}_{1}=-\vec{n}_{2}$
(topologically equivalent to one more $S^{2}$) removed. The resulting
target space $G$ is topologically equivalent
to $S^{2}$. To prove it one can show \footnote{The target space $G$ 
is, in fact, homotopic to a tangent bundle to $S^{2}$ which is in 
turn can be contracted onto $S^{2}$.}  
that it can be contracted 
along itself onto $S^{2}$. We illustrate the similar phenomenon on the 
corresponding one-dimensional example in Figure \ref{fig:cutting}.
Therefore, the relevant target space $G \sim S^{2}$. To check for 
allowed WZNW terms we calculate $H_{2}(G)=H_{2}(S^{2})=Z$. There is 
only one WZNW term allowed for the chosen target space!

%% 
 % 
 % \begin{figure}
 % \includegraphics[width=\columnwidth]{mapping}
 % \caption{\label{fig:mapping} Here we show the mapping of considered 
 % in the text. THe third dimension is taken to be ``time''. 
 % Then instead of the mapping of $S^{2}$ on to 
 % $S^{2}\times S^{2}\backslash S^{2}$ we have mapping of  
 % $S^{1}$ on to $S^{1}\times S^{1}\backslash S^{1}$. 
 % The letter is easily seen to be $S^{1}$.
 % }
 % \end{figure}
 % 
 %%

The existence of only one allowed topological term explains why only 
the full Berry phase (\ref{eq:kin}) is well defined. The most
general form (in $CP^{1}$ representation) of the topological term 
for two spins is given by
\begin{eqnarray}
     W_{E}^{top} &=&  \int_{0}^{\beta}d\tau\,\left[\rho_{1}
       z_{1}^{\dagger}\partial_{\tau}z_{1} +\rho_{2}
       z_{2}^{\dagger}\partial_{\tau}z_{2} +i\xi \dot{a} \right].
  \label{eq:tterm}
\end{eqnarray}
Coefficients $\rho_{1,2},\xi$ satisfy quantization conditions (or
gauge invariance conditions). Namely, $\rho_{1}-\xi$ and
$\rho_{2}+\xi$ must be integers. We see that for the double exchange
model of two core spins and a
single conduction electron we have (\ref{eq:kin}) which corresponds to  
particular integer values
$\rho_{1}-\xi=2S$ and $\rho_{2}+\xi=2S+1$. If by some reason the
conduction electron spends unequal time at two sites (e.g., the local
magnetic field acting on electron spin on one of sites is added) we
have an asymmetric situation $\rho_{1}\neq \rho_{2}$. However, the
combination $\rho_{1}-\xi$ and $\rho_{2}+\xi$ must stay integer to
guarantee the consistency of the theory.

%%%%%%%%%%%%%%%%%%%%%%%%%%%%%%%%%%%%%%%%%%%%%%%%%%%%%%%%%%%%%%%%%
\section{Conclusion}
    \label{sec:concl}
%%%%%%%%%%%%%%%%%%%%%%%%%%%%%%%%%%%%%%%%%%%%%%%%%%%%%%%%%%%%%%%%%

In this paper we considered the infinite Hund's coupling limit of the 
double exchange model with a given  
concentration $x$ of conduction electrons. In the limit of large core 
spins $S\gg 1$ one can think of electrons and core spins as of fast 
and slow degrees of freedom respectively.
The superficial averaging over the fast motion of electrons leads
to an inconsistent theory with effective fractional spins. Using the example of
the double exchange model on two lattice sites we showed that in addition
to the change of the value of the effective core spin on site from 
$S$ to $S+x/2$ 
a new topological term arises in the effective action. 
With this term the exponent of the 
effective action becomes single-valued and well defined. 
We showed that the full Berry phase (\ref{eq:kin}) reflects the 
change of topology of the system which occurs due to the adiabaticity 
condition --- the condition justifying the averaging over the charge 
degrees of freedom. The generalization of our results to the double 
exchange model for many spins is relatively straightforward. The only 
major difference with the case of two spins is that the average 
occupation number of a given lattice site by conduction electrons
depends on the 
configuration of core spins. This makes the magnitude of a 
``fractional'' spin on 
the given site a fluctuating quantity with correspondent 
modifications of Berry phase terms. The latter effect is not expected to 
be strong in the ferromagnetically ordered state.

\begin{acknowledgments}
We are grateful to Boris Spivak who stimulated us to think about
fractional spins in ferromagnetic metals.
The discussion of topological aspects of Berry phase with Dmitri Ivanov 
is deeply appreciated. A.~G.~A. is Alfred P.~Sloan Research Fellow.   
The work of Ar. A. is supported by DR Project 200153 at Los Alamos
National Laboratory.
\end{acknowledgments}

\appendix

\section{Path integral for single spin and fermionic
       determinant}
    \label{app:pathint}

In this section for the sake of completeness
we rederive the well-known path integral
representation for the quantum mechanics of a single
spin. We reduce the problem to a calculation
of a  simple fermionic determinant.  The method
we use differs from the commonly used
coherent state representation (see e.g., \cite{Fradkin-1991})
for the path integral.  It is
conceptually simpler but involves the gradient expansion
calculation of a fermionic determinant.  It turns out that
the similar determinant is needed for our
analysis of a double exchange model for two core spins.

\subsection{Path integral representation for spin $S$}

It is easy to check
that the representation $\vec{S}
=\frac{1}{2}\psi^{\dagger}_{\alpha}
\vec{\sigma}_{\alpha\beta}\psi_{\beta}$ with constraint
$\psi^{\dagger}_{\alpha}\psi_{\alpha}=1$ is a faithful
representation of (\ref{eq:su2}) with $S=1/2$.  Here
$\vec{\sigma}=(\sigma^{x},\sigma^{y},\sigma^{z})$ is a set
of Pauli matrices and $\psi_{\alpha},\psi^{\dagger}_{\beta}$
with $\alpha,\beta=1,2$ are annihilation (creation) Fermi
operators respectively which satisfy the following
(anti)commutation relations $\{\psi_{\alpha},\psi_{\beta}\}
=\{\psi^{\dagger}_{\alpha},\psi^{\dagger}_{\beta}\}=0$ and
$\{\psi_{\alpha},\psi^{\dagger}_{\beta}\}
=\delta_{\alpha\beta}$. In fact, this representation reflects how
spin operators appear in Nature with $\psi$ being
an annihilation operator of the physical electron.

We consider the following Hamiltonian $H=-2\Delta
\vec{n} \vec{S}$, where $\vec{S} =\frac{1}{2}
\psi^{\dagger}\vec{\sigma}\psi$ is a spin of a
fermion and
$\Delta$ is a strength of the coupling of this spin with the unit
vector $\vec{n}$.  The correspondent action is
\begin{equation}
    S =\int dt\, \psi^{\dagger}
          \left[i\partial_{t}+\Delta\hat{n}\right]\psi,
 \label{eq:actionspin}
\end{equation} 
where we introduced a notation
$\hat{n}=\vec{n}\vec{\sigma}$. 
One can think of $2\Delta \vec{n}$ as of
the magnetic field acting on the fermionic spin $\vec{S}$.
In the limit of an infinite coupling $\Delta\to \infty$
(infinite magnetic field) the fermionic spin $\vec{S}$ will
be ``frozen'' along the classical vector $\vec{n}$ so that
$\vec{n}$ in the limit of infinite $\Delta$ becomes a
direction of quantization axis of a quantum spin of a fermion.
If now we change the direction of the classical vector
$\vec{n}$ with time, the fermionic spin will follow the
direction of $\vec{n}$. Therefore, we expect that in the
limit of an infinite $\Delta$ an
effective dynamics of classical vector $\vec{n}$ induced by
a fermion will be that of the quantum spin $1/2$.

The path integral representation for spin $1/2$ is then
given by
\begin{eqnarray}
          Z &=& \int D\vec{n}\;  e^{iW[\vec{n}]},
     \label{eq:pathint}
\end{eqnarray}
where $W[\vec{n}]=\lim_{\Delta\to\infty}W^{eff}[\vec{n}]$
is a classical action corresponding to a quantum spin $1/2$.
This classical action is a functional of the trajectory of
the time dependent unit vector $\vec{n}(t)$. The
effective action $W_{eff}[\vec{n}]$ is defined as
\begin{eqnarray}
          e^{iW^{eff}[\vec{n}]} &=& \int D\psi D\psi^{\dagger} \;
          e^{i\int dt\, \psi^{\dagger}
          \left[i\partial_{t}+\Delta\hat{n}\right]\psi}.
     \label{eq:fermint}
\end{eqnarray}
The functional
integration in (\ref{eq:fermint})
is performed over Grassman variables\cite{NO-1998}
$\psi,\psi^{\dagger}$ -- classical analogs of fermionic
annihilation (creation) operators.
Integrating over fermions we obtain
\begin{eqnarray}
          W^{eff}[\vec{n}] &=& -i \ln \,{\rm Det}\,
          \left[i\partial_{t}+\Delta\hat{n}\right].
     \label{eq:effdet}
\end{eqnarray}

To generalize the result to an
arbitrary spin $S$ we can consider the same fermionic theory
(\ref{eq:fermint}) but with $2S$ independent species of
fermions coupled to the same vector $\vec{n}$.  In
this case all spins $1/2$ of fermions will be aligned along
$\vec{n}$ so that (\ref{eq:pathint}) will correspond to
quantum spin $S$.  Integration over all fermionic species
is independent and we have for an effective action
\begin{eqnarray}
          W^{eff}[\vec{n}] &=& -i 2S \ln\,{\rm Det}\,
          \left[i\partial_{t}+\Delta\hat{n}\right].
     \label{eq:effaction}
\end{eqnarray}
The only problem which is left is technical.  It is to
calculate the fermionic determinant (\ref{eq:effaction}) in the
limit of an infinite $\Delta$.

\subsection{Shortcut}

One can obtain the result without (almost) any calculations. Let us 
introduce the complex, two-component, normalized vector 
$z^{t}=(z_{1},z_{2})$, $z^{\dagger}z=1$ so that it is an eigenvector 
of $\hat{n}$ with the eigenvalue $1$: $\hat{n}z=z$. For a time 
independent $\hat{n}$ this vector is a ground state of our 
Hamiltonian $Hz=-\Delta\hat{n}z=-\Delta z$ with the ground state 
energy $-\Delta$. The other eigenstate has an energy $+\Delta$ and in 
the limit of an infinite $\Delta$ is never excited by a motion of the 
unit vector $\vec{n}$. Therefore, in the limit of an infinite $\Delta$ 
we can use $\psi_{\alpha} =z_{\alpha}\chi$ with $\chi$ -- 
spinless (projected) fermion. Substituting this into the action 
(\ref{eq:actionspin}) we obtain 
$$ S =\int dt\,\left[\chi^{\dagger}(i\partial_{t}+\Delta)\chi 
+\chi^{\dagger}\chi(z^{\dagger}i\partial_{t}z)\right].$$
The variable $\chi$ does not have any dynamics in the limit of an 
infinite $\Delta$. We average over $\chi$ using $\chi^{\dagger}\chi=1$ 
and obtain $W=\int dt\, z^{\dagger}i\partial_{t}z$ for spin $1/2$. In 
case of spin $S$ one has $2S$ species of fermions, 
$\chi^{\dagger}\chi=2S$ and $W=2S\int dt\, 
z^{\dagger}i\partial_{t}z$. This is the desired result for the 
action for a single spin $S$. However, the main point of this paper 
is to show that the ``naive'' averaging over fermions instead of the
careful determinant calculation can result in error. In this case one 
can show that due to topological reasons the calculation we performed 
is a legitimate one. To add even more weight to this statement we 
proceed with a formal calculation of the fermionic determinant 
(\ref{eq:effaction}) and show that it gives the same result for a 
single spin action.

\subsection{Determinant calculation}

Before going to the calculation of fermionic determinant
(\ref{eq:effaction}) we will perform Wick rotation of time
and go to an imaginary time representation. Imaginary time
representation is slightly more convenient for calculations
for reasons that will be clear later. The change to an
imaginary time changes quantum amplitude $e^{iW}$ into a
``Boltzmann'' factor $e^{-W_{E}}$, where subscript ``E''
stands for ``Euclidean''. In imaginary time $\tau=it$ we
have instead of (\ref{eq:effaction})
\begin{eqnarray}
          W^{eff}_{E}[\vec{n}] &=& - 2S \ln\,{\rm Det}\,
          \left[\partial_{\tau}-\Delta\hat{n}\right]
     \label{eq:effactionE}
\end{eqnarray}
with the partition function for a single spin
\begin{eqnarray}
          Z &=& \int D\vec{n}\;  e^{-W_{E}[\vec{n}]},
     \label{eq:pathintEapp}
\end{eqnarray}
where $W_{E}[\vec{n}]=\lim_{\Delta\to\infty}W_{E}^{eff}[\vec{n}]$.
We impose periodic boundary conditions (in imaginary time)
on $\vec{n}(\tau)$ so that $\vec{n}(\beta)=\vec{n}(0)$ and
$\beta$ is a maximal span of an imaginary time.  With these
boundary conditions we can think of time as of the
one-dimensional circle with identification of points
$\tau=\beta$ with $\tau=0$.  Physical results will be
obtained in the limit $\beta\to \infty$ after analytical
continuation back to real time $t=-i\tau$.

Using the method presented in  Appendix \ref{app:diagrams} one can
calculate the effective action (\ref{eq:effactionE}) in adiabatic
approximation with an arbitrary accuracy. However,  
we are interested in the limit $\Delta\to \infty$.
In this limit all terms of gradient expansion in $1/\Delta$ except for
the first (topological) one will vanish.
Here we show how to obtain the most important
(topological) $\Delta $-independent part of the effective action
(\ref{eq:effactionE}) which is the exact result
for the action for a single spin.

Determinant of an operator does not change if one applies a
unitary transformation to it. This is, however, a potentially dangerous
procedure  in the context of gradient expansion
as one has to make sure the boundary conditions
for electrons do not change in the process of such transformation.
It turns out that this is not an issue for the single spin problem.
However, e.g., for the double exchange problem 
the boundary condition of
fermions do change in a non-trivial way under the unitary rotation 
(see section \ref{sec:de2}).

Let us use the transformation
$\hat{U}$ such that $\hat{U}^{-1}\hat{n}\hat{U}=\sigma
^{3}$.  This transformation $\hat{U}$ is easily found using
the so called $CP^{1}$ representation.  In this
representation we introduce a two-component complex vector
$z$ with the constraint $z^{\dagger }z = 1$ such that
$\vec{n} = z^{\dagger }\vec{\sigma }z$.  Then the matrix
$\hat{U}$ can be written as
\begin{eqnarray}
      \hat{U} = \left(
       \begin{array}{cc}
          z_{1} & z_{2}^{\ast}  \\
          z_{2} & -z_{1}^{\ast}
       \end{array}
      \right).
  \label{eq:U}
\end{eqnarray}
We write the equation (\ref{eq:effactionE}) in the form
\begin{eqnarray}
          W_{E}^{eff}[\vec{n}] &=& - 2S \,{\rm{Tr}\,}\ln
          \left[\partial_{\tau}+\hat{A}-\Delta\sigma ^{3}\right],
     \label{eq:effactionE2}
\end{eqnarray}
where $\hat{A} = \hat{U}^{-1}\partial _{\tau }\hat{U}$ and
$\,{\rm{Tr}\,}$ means both functional trace and the trace over
$\sigma$-matrices.

The matrix elements of $\hat{A}$ are proportional to the first time
derivative of spin variables and are small in comparison to $\Delta$.
We can expand the logarithm in (\ref{eq:effactionE2}) 
in powers
of $\hat{A}$. We keep only the first $\hat{A}$ dependent term
as the other terms will have additional powers of $\Delta$ in 
denominators. (one can obtain those terms in a regular way explained 
in Appendix \ref{app:diagrams})
\begin{eqnarray}
     W_{E}^{eff}[\vec{n}] &=& 2S \,{\rm{Tr}\,}\frac{\partial _{\tau }
     +\Delta \sigma ^{3} }
     {-\partial ^{2}_{\tau }+\Delta ^{2}}\hat{A}
  \nonumber \\
     &=& i\int \frac{d\omega }{2\pi }
     \,\mbox{tr}\,\frac{-\omega +i\Delta \sigma ^{3}}
     {\omega ^{2}+\Delta ^{2}}\hat{A}(\omega =0)
\nonumber \\
     &=& 2S\frac{1}{2}\int_{0}^{\beta } d\tau \,\mbox{tr}\,
     \sigma ^{3}\hat{A}
  \nonumber \\
     &=& - i 2S\int_{0}^{\beta } d\tau\, z^{\dagger }i\partial _{\tau }z.
     \label{eq:effactionE3}
\end{eqnarray}

It is instructive to rewrite (\ref{eq:effactionE3}) in terms of 
the original variables $\vec{n}$.
We parameterize complex vector
$z^{t}=e^{i\psi}(\cos\frac{\theta}{2}, e^{i\phi}\sin\frac{\theta}{2})
$. It is easy to see that the angles $\theta$ and $\phi$ are the polar
and azimuthal angles corresponding to the direction of the unit vector
$\vec{n}$. The phase $\psi$ can be chosen arbitrarily as it does not 
affect $\vec{n}$. Up to the 
full time derivative of $\psi$ we obtain  
$-\int_{0}^{\beta } d\tau\, z^{\dagger }i\partial _{\tau }z =
\frac{1}{2}\int_{0}^{\beta } d\tau\, (1-\cos\theta)\dot{\phi}=\Omega/2$,
where $\Omega$ is a solid angle enclosed by the unit vector $\vec{n}$
during its time evolution. One can check that this solid angle 
can be rewritten in the form (\ref{eq:WZNW}) which is 
explicitly rotationally invariant.

%%%%%%%%%%%%%%%%%%%%%%%%%%%%%%%%%%%%%%%%%%%%%%%%%%%%%%%%%%%%%%%%%%%%%%%%
\section{Calculation of fermionic determinant for double exchange
model}
   \label{app:diagrams}
%%%%%%%%%%%%%%%%%%%%%%%%%%%%%%%%%%%%%%%%%%%%%%%%%%%%%%%%%%%%%%%%%%%%%%%%
                                                                         %
In this appendix we calculate the effective action (\ref{eq:ft2})
which can be written as
\[
W_{E}^{ind}=-\,{\rm{Tr}\,}\ln D,
\]
where the differential operator $D$ is defined as
\[
D = \partial_{\tau}+\hat{A}
\]
with matrix $\hat{A}$ given by (\ref{eq:A2},\ref{eq:A20}). We rewrite
\[
\hat{A} = iA_{0}/2+
ia_{0}\sigma ^{3}+\Delta_{1}\sigma ^{1}+\Delta _{2}\sigma ^{2} =
iA_{0}/2+\vec{\Delta}\vec{\sigma },
\]
where $\Delta_{1}=-\Delta\cos a $, $\Delta_{2}=-\Delta\sin a $,
and $\Delta_{3}=ia_{0}$.

We use an adiabatic approximation using the (gradient) expansion in
$1/\Delta$.
The operator $D$ contains the magnitude of the effective hopping
$\Delta$ which is slowly varying but it is not small itself. To avoid
this difficulty we calculate instead of effective action its variation
and then restore the result from this variation
\begin{equation}
     \delta W_{E}^{ind} = -\,{\rm{Tr}\,}\delta D D^{-1} = -\,{\rm{Tr}\,}\left[
     \delta\hat{A}\frac{1}{\partial_{\tau} +\hat{A}}\right],
  \nonumber
\end{equation}
where
\[
\delta \hat{A} = i\delta A_{0}/2 + \delta \vec{\Delta} \vec{\sigma }.
\]
Let us now calculate the functional trace in the basis of plane waves
\begin{equation}
     \delta W_{E}^{ind} = -\,{\rm{tr}\,}\int\frac{d\omega }{2\pi }\,\int
     d\tau\,
      e^{-i\omega \tau}\delta \hat{A}\frac{1}{\partial_{\tau} 
+\hat{A}}e^{i\omega
      \tau}
\end{equation}
Here the integral over time $\tau$ is the calculation of the diagonal
matrix element $\langle \omega| \delta D D^{-1} |\omega\rangle$ and
integration over $\omega$ is the calculation of the functional trace.
The residual trace $\,{\rm{tr}\,}$ is the trace in the space of $2\times 2$
matrices.
``Pulling'' $e^{i\omega\tau}$ from right to the left we have
\begin{equation}
     \delta W_{E}^{ind} = -\,{\rm{tr}\,}\int\frac{d\omega }{2\pi }\,\int
     d\tau\,
      \delta \hat{A}\frac{1}{i\omega+\partial_{\tau} +\hat{A}}.
\end{equation}
In the latter expression the partial derivative $\partial_{\tau}$ is
assumed to act ``to the right''. This means that when we expand
\begin{eqnarray}
      \delta W_{E}^{ind} &=& -{\rm tr}\int dt\int\frac{d\omega }{2\pi }
      \delta \hat{A}G\left[1 - \partial_{\tau} G
      +\partial_{\tau} G \partial_{\tau} G + \dots \right]
  \nonumber \\
        &=&
      \int dt\left[\delta L_{0}+\delta L_{1}+\delta L_{2}+\dots
      \right],
  \label{eq:expansion}
\end{eqnarray}
all derivatives are assumed to act on all terms to the right. E.g.,
$\partial_{\tau} G \partial_{\tau} G = G \ddot{G} +\dot{G}\dot{G}$.
We introduced
\begin{equation}
  \label{Green}
     G = \frac{1}{i\omega +\hat{A}}=
     \frac{-i(\omega +A_{0}/2)+\vec{\Delta} \vec{\sigma } }
     {(\omega + A_{0}/2)^{2}+|\vec{\Delta}|^{2}}
\end{equation}
and
\begin{eqnarray}
     \delta L_{0} &=& 
     -\,{\rm{tr}\,}\int \frac{d\omega }{2\pi }\delta \hat{A}G,
  \nonumber \\
     \delta L_{1} &=& \,{\rm{tr}\,}\int \frac{d\omega }{2\pi }
     \delta \hat{A}G\partial_{\tau} G,
  \label{deltaL} \\
     \delta L_{2} &=& -\,{\rm{tr}\,}\int \frac{d\omega }{2\pi }
     \delta \hat{A}G\partial_{\tau} G\partial_{\tau} G  .
  \nonumber
\end{eqnarray}
We notice that $|\vec{\Delta}|^{2}=\Delta^{2}-a_{0}^{2}\approx \Delta^{2}$
because $a_{0}\ll \Delta$ by adiabatic approximation (it contains time
derivative of core spins). Therefore, the denominator of (\ref{Green})
is never zero and integrals (\ref{deltaL}) are not singular.
The expansion (\ref{eq:expansion}) is essentially the gradient
expansion, i.e., the expansion in the number of time derivatives. One
can see that $L_{1}$ contains at least one time derivative,
$L_{2}$ -- at least two etc.

The calculation of $\delta L_{0}$ is straightforward and gives
\begin{equation}
     \delta L_{0}=\delta \left[iA_{0}/2-|\vec{\Delta}|  \right] 
  \label{deltaL0}
\end{equation}
or removing variation
\begin{equation}
     L_{0}=i\frac{A_{0}}{2}-|\vec{\Delta}|  
     \approx i\frac{A_{0}}{2}-\Delta +\frac{a_{0}^{2}}{2\Delta}. 
  \label{fullL0}
\end{equation}
To obtain the latter expression we expanded up to the second order in 
time derivatives using $|\vec{\Delta}| =\sqrt{\Delta^{2}-a_{0}^{2}}
\approx \Delta -a_{0}^{2}/2\Delta$.

Let us now prove that the whole dependence of the full result
(\ref{eq:expansion}) on $A_{0}$ comes from $L_{0}$.
Varying the action (\ref{eq:proj2}) over $A_{0}$ we obtain
$\frac{i}{2}\langle\Psi^{\dagger}\Psi\rangle = \frac{i}{2}$. This
statement is exact because the total number of electrons in the
system is fixed $\langle\Psi^{\dagger}\Psi\rangle =1$. For the
effective action we have, therefore, $\frac{\delta W_{E}^{ind}}{\delta
A_{0}}=\frac{i}{2}$. However, from (\ref{deltaL0}) we have
$\frac{\delta L_{0}}{\delta
A_{0}}=\frac{i}{2}$ which gives $\frac{\delta (W_{E}^{ind}-L_{0})}{\delta
A_{0}}=0$ and all higher order terms $L_{1}$, $L_{2}$ etc. do not
depend on $A_{0}$. This simplifies further calculations allowing to
use
$$G=\left.\frac{1}{i\omega +\hat{A}}\right|_{A_{0}=0}=
     \frac{-i\omega+\vec{\Delta} \vec{\sigma } }
     {\omega^{2}+|\vec{\Delta}|^{2}}
$$
for all higher order terms.

As a result of calculations one obtains from (\ref{deltaL})
\begin{equation}
    \delta L_{1} =\frac{i}{2}
    \epsilon^{abc}
    \left(\delta \frac{\Delta^{a}}{|\vec{\Delta}|}\right)
    \frac{\Delta^{b}}{|\vec{\Delta}|}
    \partial_{\tau}\frac{\Delta^{c}}{|\vec{\Delta}|}
 \label{eq:L1}
\end{equation}
This is identical in form to a Berry phase of a single spin $1/2$ 
given by (\ref{eq:variation}) with identification 
$\vec{n}=\vec{\Delta}/|\vec{\Delta}|$. Therefore, $L_{1}$ is given by 
formula analogous to the Berry phase for a single spin (\ref{eq:WZNW}) 
and is to be interpreted as a half of solid angle swept by vector 
$\vec{\Delta}$ during its time evolution. Assume for a moment now 
that $a_{0}=0$. Then the vector $\vec{\Delta}/|\vec{\Delta}|
=(-\cos a,-\sin a,0)$ 
moves along the equator of unit sphere and covers the solid angle 
$\Omega = a(\beta)-a(0)$ during its motion. In this approximation we 
obtain $L_{1}=i\dot{a}/2$ which gives half of the solid angle after 
time integration \footnote{The method we used to calculate the 
geometric phase $\dot{a}/2$ is known as the method of embedding in 
quantum field theory \cite{Witten-1983global}}. 
Calculating corrections due to nonzero $a_{0}$ 
from  (\ref{eq:L1}) we obtain
\begin{equation}
      L_{1} = \frac{i}{2} \left(\dot{a} 
      -\frac{ia_{0}\dot{a}}{|\vec{\Delta}|}\right)
      \approx \frac{i}{2} \dot{a} 
      -\frac{a_{0}\dot{a}}{2\Delta},
 \label{fullL1}  
\end{equation}
where we expanded the latter expression up to the terms 
of the second order in 
time derivatives.

The calculation of $L_{2}$ in (\ref{deltaL}) is straightforward and 
gives
\begin{equation}
      L_{2} =
      \frac{1}{8|\vec{\Delta}|}
      \left(\partial_{\tau} \frac{\vec{\Delta}}{|\vec{\Delta}|} 
      \right)^{2}
      \approx \frac{\dot{a}^{2}}{8\Delta}.
   \label{fullL2}
\end{equation}

Finally, combining terms (\ref{fullL0},\ref{fullL1},\ref{fullL2}) we 
obtain an effective action up to the second order terms
\begin{equation}
      L_{E} =\frac{i}{2}\left(A_{0}+\dot{a}\right)
      -\Delta +\frac{1}{2\Delta}\left(a_{0}-\frac{\dot{a}}{2} 
      \right)^{2}.
   \label{fullLE}
\end{equation}
It is interesting to notice that although neither of contributions 
(\ref{fullL0},\ref{fullL1},\ref{fullL2}) is gauge invariant, the full 
action (\ref{fullLE}) is gauge invariant. This means that the final 
result (\ref{fullLE}) can be rewritten in terms of two unit vectors 
$\vec{n}_{1}$ and $\vec{n}_{2}$ which represent physical degrees of 
freedom -- core spins of double exchange model (\ref{eq:de}).

The method of gradient expansion used in this appendix can of course 
also give higher order corrections to (\ref{fullLE}).

%%%%%%%%%%%%%%%%%%%%%%%%%%%%%%%%%%%%%%%%%%%%%%%%%%%%%%%%%%%%%%%%%%%%%
%%%%%%%%%%%%%%%%%%%%%%%%%%%%%%%%%%%%%%%%%%%%%%%%%%%%%%%%%%%%%%%%%%%%%
%%%%%%%%%%%%%%%%%%%%%%%%%%%%%%%%%%%%%%%%%%%%%%%%%%%%%%%%%%%%%%%%%%%%%
%%%%%%%%%%%%%%%%%%%%%%%%%%%%%%%%%%%%%%%%%%%%%%%%%%%%%%%%%%%%%%%%%%%%%
%%%%%%%%%%%%%%%%%%%%%%%%%%%%%%%%%%%%%%%%%%%%%%%%%%%%%%%%%%%%%%%%%%%%%
                                                                    % 

%\end{multicols}

\begin{thebibliography}{99}
    
    

\bibitem{Zener-1951}
C. Zener, Phys. Rev. {\bf 82}, 403 (1951).
\\ {\it Interaction between the d-shells in the transition
metals. 2. Ferromagnetic compounds of manganese with
perovskite structure.}

\bibitem{AH-1955}
P. W. Anderson and H. Hasegawa,
Phys. Rev. {\bf 100}, 675 (1955).
\\ {\it Considerations on double exchange.}

\bibitem{double_exchange}
The limit of $J_{H}\gg t$, $S\gg 1$ of double exchange model can be a 
good starting points to study 
the physics of manganite perovskites \cite{MLS-1995}. For experimental 
review see \cite{Ramirez-1997}.

\bibitem{MLS-1995}
A.J. Millis, P.B. Littlewood, and B.I. Shraiman, Phys. Rev. Lett. 
{\bf 74}, 5144 (1995).
\\ {\it Double Exchange Alone Does Not Explain the Resistivity of 
$La_{1-x}Sr_{x}MnO_{3}$.} 

\bibitem{Ramirez-1997}
A.~P. Ramirez, J. Phys.: Condens. Matter {\bf 9}, 8171 (1997).
\\ {\it Colossal magnetoresistance.}

\bibitem{deGennes-1960}
P. -G. de Gennes,
Phys. Rev. 118, 141-154 (1960).
\\ {\it Effects of Double Exchange in Magnetic Crystals.}

\bibitem{Fradkin-1991}
Eduardo Fradkin,
{\it Field Theories of Condensed Matter Systems},
Redwood City, Calif.:
Addison-Wesley Pub. Co., 1991.

\bibitem{Berry-1984} 
M. Berry, Proc. R. Soc. Lond. A 392, 45 (1984).

\bibitem{currentalgebras}
S.~B.~Treiman, R.~Jackiw, B.~Zumino, and E.~Witten,
{\it Current Algebra and Anomalies},
(Princeton University Press, 1985).

\bibitem{NO-1998} 
As a general reference on integration over Grassman variables see, 
e.g., 
J.~W.~Negele and H.~Orland,
{\it Quantum Many-Particle Systems},
Perseus Pr, 1998.

\bibitem{Witten-1983global}
E.~Witten, Nucl. Phys. {\bf B} {\bf 223}, 422 (1983).
\\ {\it Global aspects of current algebra.}



\end{thebibliography}
\end{document}